\documentclass[11pt,a4paper]{article}
\usepackage{amsmath}

\usepackage[psamsfonts]{amssymb}
\usepackage{amsmath}
\usepackage{epsfig}
\usepackage{float}
\author{H. Mohseni Sadjadi \footnote{mohsenisad@ut.ac.ir} and Parviz Goodarzi
\\ {\small Department of Physics, University of Tehran,}
\\ {\small P. O. B. 14395-547, Tehran 14399-55961, Iran}}
\title{Oscillatory inflation in non-minimal derivative coupling model}

\begin{document}
\maketitle
\begin{abstract}
Inflation during rapid oscillation of a scalar field in
non-minimal derivative coupling model is discussed. Cosmological
perturbations originated in this stage are studied and the
consistency of the results with observational constraints coming
from Planck 2013 data are investigated.
\end{abstract}

\section{Introduction}
In the past three decades various models have been proposed for
inflation \cite{guth}, where in many of them inflation is driven by
a canonical scalar field, $\phi$ (dubbed inflaton), rolling slowly
in an almost flat potential. Higgs boson may be a natural candidate
for inflaton \cite{berz}. Inspired by this idea, the authors of
\cite{Germani1}, by introducing a non-minimal coupling between
kinetic term of the scalar field and the Einstein tensor, tried to
consider the inflaton as the Higgs boson, without violating the
unitarity bound. This model is specified by the action
\begin{equation}\label{1}
S=\int \Big({M_P^2\over 2}R-{1\over 2}\Delta^{\mu \nu}\partial_\mu
\phi \partial_{\nu} \phi- V(\phi)\Big) \sqrt{-g}d^4x,
\end{equation}
where $\Delta^{\mu \nu}=g^{\mu \nu}-{1\over M^2}G^{\mu \nu}$, and
$G^{\mu \nu}=R^{\mu \nu}-{1\over 2}Rg^{\mu \nu}$ is the Einstein
tensor. The minus sign before the Einstein tensor prevents ghost presence in the theory.
$M$ is a coupling constant with the dimension of mass, and
$M_P=2.435\times 10^{18}GeV$ is the reduced Planck mass. Inflation
\cite{Germani1}, rapid oscillation \cite{Sadj1}, reheating
\cite{Sadj2}, and late time acceleration \cite{Sadj3}, have been
recently studied in the context of this non minimal derivative
coupling model. To find more features of this model see
\cite{infl}.

In the literature, it is often assumed that inflation nearly
ceases after the slow-roll and the inflaton enters a rapid
oscillation phase during which the radiation is generated. But in
\cite{Mukh} the possibility of continuation of inflation during
rapid oscillation phase for a potential satisfying a non-convexity
inequality was proposed. Inflation continues as long as the scalar
field is trapped in the convex core. This effect was reported and
numerically confirmed in \cite{Liddle}, where it was shown that
only a few number of e-folds is realized in this era. Due to the
few number of e-folds in the rapid oscillation phase, it is expected that cosmological
perturbations, as the seed of structure formation, were originated
in the slow-roll regime. However perturbations originated in the
rapid oscillation era may have imprints on cosmological scales
provided that one considers an adequate period of inflation during
rapid oscillation. This may happen in more complicated models such
as hybrid inflation, as was asserted in \cite{Mukh}.

In this work we study inflation during rapid oscillation in
non-minimal derivative coupling model proposed in \cite{Germani1}.
Conditions required for this oscillation and also inflation in
this stage are discussed and the possibility that the inflation
ceases is studied.  Cosmological perturbations created in this era
are computed and the consistency of the results with observational
constraints coming from Planck 2013 \cite{Planck1} are
investigated.

\section{Oscillatory inflation}

We consider gravitational enhanced friction  model (\ref{1}) in
the spatially flat Friedmann-Lema\^{\i}tre-Robertson-Walker (FLRW)
space-time.  The scalar field equation of motion is
\begin{equation}\label{3.5}
(1+{3H^2\over M^2})\ddot{\phi}+3H(1+{3H^2\over M^2}+{2\dot{H}\over
M^2})\dot{\phi}+V'(\phi)=0,
\end{equation}
where $H={\dot{a}\over a}$ is the Hubble parameter and a dot is
the differentiation with respect to the cosmic time $t$. The
energy density and the pressure for this homogeneous and isotopic
scalar field can be expressed as
\begin{equation}\label{2}
\rho_\phi=(1+{9H^2\over M^2}){\dot{\phi}^2\over 2}+V(\phi),
\end{equation}
and
\begin{equation}\label{3}
P_\phi=(1-{3H^2\over M^2}-{2\dot{H}\over M^2}){\dot{\phi}^2\over
2}-V(\phi)-{2H\dot{\phi}\ddot{\phi}\over M^2},
\end{equation}
respectively. The Friedmann equation reads
\begin{equation}\label{s1}
H^2={1\over 3M_P^2}\rho_\phi.
\end{equation}

\begin{figure}[h]
\centering\epsfig{file=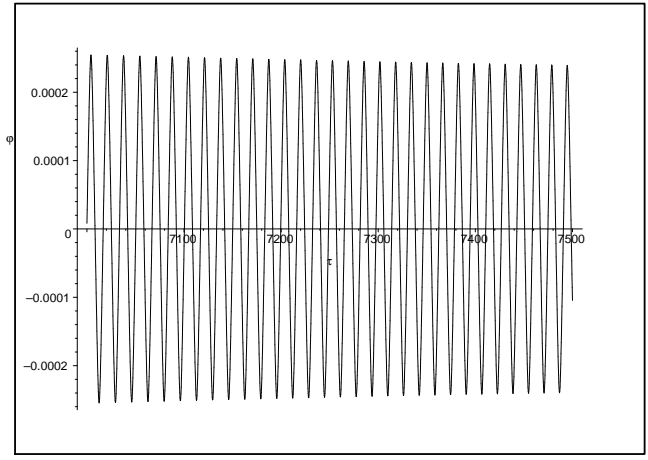,width=6cm,angle=270}
\caption{${\varphi:={\phi\over M_P}}$ in terms of dimensionless time
$\tau=mt$, for ${m^2\over M^2}=10^{8}$ with $\{\varphi(1)=0.01$,
$\dot{\varphi}(1)=0\}$, for the quadratic potential, ${1\over
2}m^2\phi^2$. } \label{fig0}
\end{figure}

\begin{figure}[!]
\centering\epsfig{file=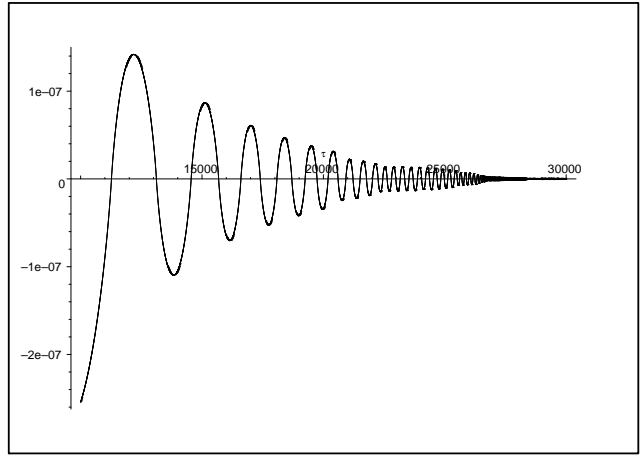,width=6cm,angle=270}
\caption{${\varphi:={\phi\over M_P}}$ in terms of dimensionless time
$\tau=M_Pt$, for  $M=10^{-9}M_P$ and $\lambda=1.76\times 10^{-8}M_P^{4-0.0392}$
with initial conditions $\{\varphi(1)=10^{-6}$,
$\dot{\varphi}(1)=0\}$. } \label{fig00}
\end{figure}

The slow roll solution and the associated inflation were studied in
\cite{Germani1}. Here we consider  rapid oscillatory solution for
the scalar field, with time dependent amplitude $\Phi(t)$ (the
highest point of oscillation at which $\dot{\phi}=0$) and also time
dependent frequency $\omega(t)={1\over T(t)}$. $T(t)$ is the period
of the oscillation
\begin{equation}\label{s3}
T=2\int_{-\Phi}^{\Phi}{d\phi\over \dot{\phi}}.
\end{equation}
The rapid oscillation phase is characterized by
\begin{equation}\label{4}
H(t)\ll {1\over T(t)}
\end{equation}
and
\begin{equation}\label{ref1}
\left|{\dot{H}\over H}\right|\ll {1\over T},
\end{equation}
implying that the Hubble parameter is much smaller than the time
dependent frequency and changes insignificantly during one
oscillation: $H(t')\approx H(t)$ for $t\leq t'\leq t+T(t)$. From
(\ref{s1}) and (\ref{ref1}), it is clear that like $H$,
$\rho_{\phi}$ remains approximately a constant during one period. We
take this nearly constant as the value of the energy density at the
amplitude, $\Phi$, where $\dot{\phi}\big|_{|\phi|=\Phi}=0$ (see fig.
(\ref{fig0})). Therefore the energy density during one oscillation
can be expressed in terms of the value of the potential at the
corresponding amplitude \cite{rapid},
\begin{equation}\label{s4}
\rho_{\phi}\approx V(\Phi).
\end{equation}
Therefore for a power law potential we expect that
$\left|{\dot\Phi\over \Phi}\right|\ll {1\over T}$. To elucidate more
this subject, in fig.(\ref{fig0}), the rapid oscillating scalar
field is depicted numerically by using eqs. (\ref{2}), (\ref{3.5}),
and (\ref{s1}) for a quadratic potential, showing that the amplitude
of oscillation changes very slowly during one oscillation.
A more detailed discussion about this solution may be found in
\cite{rapid} and \cite{Sadj1}.
Also, in fig.(\ref{fig00}) the oscillation of the scalar field
for the potential $V(\phi)=\lambda \left|\phi \right|^{0.0392}$ is numerically shown  (the reason for this choice will be revealed when
we will determine our parameters from astrophysical data in the third section).

The adiabatic index of the scalar field, defined by $\gamma=w+1$
where $w$ is the equation of state parameter (EoS): $w={P_\phi\over
\rho_\phi}$, in the rapid oscillation phase is effectively given by
\begin{eqnarray}\label{5}
\gamma&=&\left<P_\phi+\rho_\phi\right>\over
\left<\rho_\phi\right>\nonumber \\
&=&{\big<{\left(1+{3H^2\over
M^2}\right)\dot{\phi}^2-{d{({H\dot{\phi}^2\over M^2})}\over
dt}}\big>\over<\rho_\phi>}\nonumber \\
&=& \left(1+{3H^2\over
M^2}\right){\big<\dot{\phi}^2\big>\over\big<\rho_\phi\big>}\nonumber \\
&=&{2\left(1+{3H^2\over M^2}\right)\over \left(1+{9H^2\over
M^2}\right)}{\left<\rho_\phi-V(\phi)\right>\over
\left<\rho_\phi\right>}\nonumber \\
&=&{2\left(1+{3H^2\over M^2}\right)\over \left(1+{9H^2\over
M^2}\right)V(\Phi)}{\int_{-\Phi}^{\Phi}\sqrt{V(\Phi)-V(\phi)}d\phi\over
\int_{-\Phi}^\Phi{d\phi\over \sqrt{V(\Phi)-V(\phi)}}}
\end{eqnarray}
$<..>={\int_t^{t+T}..dt'\over T}$ is the average over an oscillation
with period $T$. To obtain (\ref{5}), we have used (\ref{2}),
(\ref{3}) and (\ref{s4}), and have taken into the account the fact that $\dot{\phi}$ vanishes
at $\left|\phi\right|=\Phi$.  (\ref{5}) is valid only for time scale
much larger than the period $T$: $t\gg T$, over which the average
was taken. For even power law potentials
\begin{equation}\label{s7}
V(\phi)=\lambda \phi^q,
\end{equation}
where $\lambda\in \Re$, the adiabatic index becomes
\begin{equation}\label{s8}
\gamma={2q\over q+2}{1+{3H^2\over M^2}\over 1+{9H^2\over M^2}}.
\end{equation}
In the minimal coupling, $M\to \infty$, we recover the result
derived in \cite{rapid}  (see also \cite{Mukh}, \cite{Liddle},
\cite{kolb1} and references therein), $\gamma={2q\over q+2}$. We
have also
\begin{equation}\label{s5}
\left<\dot{\rho_\phi}\right>={\rho_\phi(t+T)-\rho_\phi(t)\over
T}\simeq \dot{\rho_\phi},
\end{equation}
where $t\gg T$ has been used. By taking the average of the
continuity equation, namely $\left
<\dot{\rho_\phi}+3H(P_\phi+\rho_\phi)\right>=0 $ we obtain
\begin{equation}\label{con}
\dot{\rho_\phi}+3H\gamma\rho_\phi=0,
\end{equation}
where $\gamma$ is given by (\ref{5}). All the quantities in the
above equation must be regarded as their averaged value in the
sense explained above and the equation is valid for large time
with respect to the period $T$. For a constant $\gamma$, the system composed of the equations
(\ref{con}) and (\ref{s1}) may be solved analytically. For the
power law potential, this occurs for minimal model and also when
we consider the high friction regime \cite{Germani1},
\begin{equation}\label{s2}
{H^2\over M^2}\gg 1,
\end{equation}
leading to $\gamma={2q\over 3q +6}$. In this situation analytical
solutions for the energy density, the scale factor, and the Hubble
Parameter are
\begin{equation}\label{10}
\rho_\phi\propto a^{-3\gamma},
\end{equation}
\begin{equation}\label{11}
a\propto t^{2\over 3\gamma},
\end{equation}
\begin{eqnarray}\label{s9}
H={2\over 3\gamma t},
\end{eqnarray}
respectively.  Here, in contrast to the slow roll, $\left|\dot{H}\right|$ and $H^2$ may be of the same order of magnitude. From the continuity equation one can find that the
amplitude, $\Phi$, satisfies
\begin{equation}\label{s10}
\dot{\Phi}+{3\gamma H\over q}\Phi=0
\end{equation}
whose solution is
\begin{equation}\label{13}
\Phi(t)\propto a^{{-3\gamma\over q}}\propto t^{-{2\over q}}.
\end{equation}
Hereafter we restrict ourselves to the high friction regime
(\ref{s2}), where as we have seen, analytical solution for the
problem can be found. Note that in derivation of these solutions
we have employed the conditions (\ref{s1}), (\ref{4}) and $t\gg T$. So the domain of validity of our result
is where the solutions satisfy these conditions. Using (\ref{s9}), we find that if $HT\ll 1$ is satisfied,
then (\ref{s1}) and $t\gg T$ are also fulfilled. In the case of the power law potential, the period is determined as
\begin{eqnarray}\label{s11}
T&=&2\int_{-\Phi}^{\Phi}{d\phi\over \dot{\phi}}\nonumber
\\
&=&\sqrt{{18H^2\over M^2}}\int_{-\Phi}^{\Phi}{d\phi\over
{\sqrt{\rho_\phi-V(\phi)}}}\nonumber \\
&=&\sqrt{{18H^2\over M^2}}\int_{-\Phi}^{\Phi}{d\phi\over
{\sqrt{\lambda \Phi^q-\lambda \phi^q}}}\nonumber \\
&=& 2\sqrt{18 \pi H^2\over \lambda M^2}{\Gamma\left( {1\over
q}\right)\over q\Gamma\left( {q+2\over 2q}\right)}\Phi^{{2-q}\over
2}\nonumber \\
&=&{\sqrt{24\pi}\over M_PM}{\Gamma\left({1\over q}\right)\over q\Gamma\left({q+2\over 2q}\right)}\Phi.
\end{eqnarray}
where  $H^2={\lambda\over 3M_P^2}\Phi^q$, derived from (\ref{s1}),
(\ref{s4}), and (\ref{s7}) has been used. Hence, $HT\ll 1$ can be
rewritten in terms of $\Phi$ as
\begin{equation}\label{s12}
\Phi^{q+2\over 2}\ll \left({q\Gamma\left( {q+2\over
2q}\right)\over \sqrt{8\pi} {\Gamma\left( {1\over
q}\right)}}\right){M_P^2M\over \sqrt{\lambda}}.
\end{equation}
The presence of the scale $M$ reduces the scale of the scalar
field with respect to the minimal case in which the same procedure
gives \cite{Gu}
\begin{equation}\label{s13}
\Phi\ll  \left({q\Gamma\left( {q+2\over 2q}\right)\over
\sqrt{8\pi} {\Gamma\left( {1\over q}\right)}}\right)\sqrt{3}M_P.
\end{equation}
This can also be rewritten in terms of the Hubble parameter as
\begin{equation}\label{hr}
H^{q+2\over q}\ll \left({q\Gamma\left( {q+2\over 2q}\right)\over
\sqrt{8\pi}3^{q+2\over 2q} {\Gamma\left( {1\over
q}\right)}}\right)\lambda^{1\over q}M M_P^{q-2\over q}.
\end{equation}
Therefore the domain of validity of our solutions is given by
(\ref{s12}) or (\ref{hr}) which specifies a bound for the Hubble
parameter (and consequently for the energy density) during rapid
oscillation.

Note that in the slow roll we had  $\ddot{\phi}\ll \dot 3H\dot{\phi}$,  and also $\rho_\phi\approx V(\phi)$ which together with (\ref{2}) imply
\begin{equation}\label{r1}
\left(1+9{H^2\over M^2}\right){\dot{\phi}^2\over 2}\ll V(\phi).
\end{equation}
In the high friction regime, these conditions are satisfied  (for details see \cite{Sadj1}) when $\phi^{q+2}\gg {q^2 M^2M_P^4\over \lambda}$, which is opposite to (\ref{s12}). In the high friction regime (\ref{r1}) leads to ${9H^2\over 2M^2}\dot{\phi}^2\ll V(\phi)\sim 3M_P^2H^2$ resulting
$\dot{\phi}^2\ll {2\over 3}M^2M_P^2$. In contrast to this result, in quasi periodic stage, (\ref{5}) and (\ref{s1})
result in
\begin{equation}\label{14}
\big<\dot{\phi}^2\big>\approx\gamma M_p^2 M^2.
\end{equation}

Inflation occurs when $\ddot{a}>0$ or in terms of the adiabatic
index: $\gamma<{2\over 3}$ which leads to $q\in (-2,\infty)$. Note
that in the minimal case, where $\gamma={2q\over q+2}$, inflation
takes place only for the short range $q\in (-2,1)$. Inflation
continues as long as $\gamma<{2\over 3}$, which from the fourth equality in (\ref{5})
leads to
\begin{equation}\label{infgamma}
{1+{3H^2\over M^2}\over 1+{9H^2\over M^2}}\left<\rho_\phi-V\right><{1\over 3}<\rho_\phi>
\end{equation}
In our high friction regime,  this  reduces to the simple inequality
\begin{equation}\label{s14}
\left<V(\phi)\right> >0,
\end{equation}
while in the minimal case, ${H^2\over M^2}\to 0$,
a more complicated inequality, $\left<V(\phi)\right>>{2\over 3}\left<\rho_\phi\right>$, arises, which
using $<\dot{\phi}^2>=<\phi V'(\phi)>$, leads to \cite{Mukh}: $\left<V(\phi)-\phi V'(\phi)\right>
>0$.

For simple power law potentials and in the rapid oscillation phase
$\gamma$ is a constant and consequently inflation does not cease
without taking into the account another formalism such as particle
production. Indeed if one considers interaction between the scalar field and other components
such as radiation, the energy of the scalar field
is released and, depending on the coupling, rapid oscillatory inflation may be
promptly terminated in this situation.
This possibility is discussed in \cite{ref},  where inflation and reheating are studied  in the framework of
an effective action consisting of a Galileon scalar field.

\begin{figure}[h]
\centering\epsfig{file=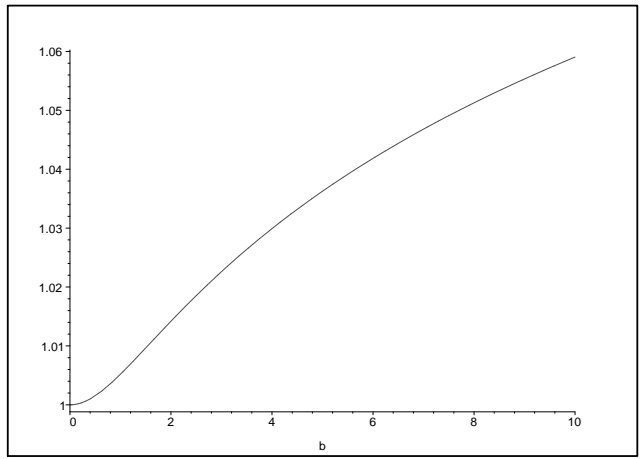,width=6cm,angle=270}
\caption{$g(b,0.0392)$ in the equation (\ref{s19}) in terms of
$b$. } \label{fig2}
\end{figure}

Inflation ends also for more complicated potential such as the
potential suggested by Damour-Mukhanov \cite{Mukh}
\begin{equation}\label{s15}
V(\phi)=v\left(\left({\phi^2\over \phi^2_c}+1\right)^{q\over
2}-d\right),
\end{equation}
where $d$ is a positive real number, $v>0$ and
$\phi_c$ are real parameters with dimension $[mass]^{4}$ and
$[mass]$ respectively (note that for large $\phi$,  $\phi\gg \phi_c$,
(\ref{s15}) reduces to a simple power law potential). To see this, we follow the same steps as
\cite{sami}. By using
\begin{equation}\label{s16}
\left<V(\phi)\right>={\int_{-\Phi}^{\Phi}{V(\phi)\over
\dot{\phi}}d\phi\over \int_{-\Phi}^{\Phi}{d\phi \over \dot{\phi}
}},
\end{equation}
we find that the inflation  continues as long as
$\int_{-\Phi}^{\Phi}V(\phi)d\phi>0$, which for the potential (\ref{s15}) gives
\begin{equation}\label{s17}
\int_{-1}^{1}\left((b^2x^2+1)^{q\over 2}-d\right)dx>0.
\end{equation}
We have defined  $b={\Phi\over \phi_c}$ and $x={\phi \over \Phi}$.
(\ref{s17}) results in that the inflation continues whenever
\begin{equation}\label{s18}
d< g(b,q),
\end{equation}
where
\begin{equation}\label{s19}
g(b,q)={G(b,q)
\over {2b}  \Gamma  \left( {q+3\over 2}
 \right) \left( 1+q \right)  \Gamma  \left( -{q\over 2},
 \right)}
\end{equation}
in which $G(b,q)=
-{\pi }^{3\over 2}(q+1)\sec \left({\pi q\over 2}\right)
+2\, \left( {b}^{2} \right) ^{{1+q\over 2}}{2F1({-{q\over
2}},-{1+q\over 2};\,{1-q\over 2};\,-{b}^{-2})}\newline
\times \Gamma
\left( -{q\over 2}\right) \Gamma  \left( {q+3\over 2} \right)$.
$2F1$ is the Gauss hypergeometric function. The inflation
ceases at $t_{end}$, i.e. when this inequality is violated such
that $d=g(b_{end},q)$ and $d>g(b(t>t_{end}),q)$. In fig.(\ref{fig2}), $g(b,q)$ in
(\ref{s19}) is numerically depicted for $q=0.0392$ (our reason for this choice will be revealed in the next section)
in terms of $b$, which shows that for $d>1$, the inflation ends for some real
value of $b$.
For $d\simeq 1$, the inflation ends for $\Phi\sim \phi_c$.

We use the same definition for the number of e-folds, from a specific time $t_{*}$ until the end of inflation, as
\cite{Liddle}
\begin{equation}\label{s20}
\mathcal{N}=\ln{a_{end}H_{end}\over a_{*}H_{*}}.
\end{equation}
In this definition $\mathcal{N}$ is a measure of $\ln(aH)$
increase during inflation. We have $\mathcal{N}=\ln{a_{end}\over
a_{*}}+\ln{H_{end}\over H_{*}}$, and as $\dot{H}<0$, $\mathcal{N}$
is less than the more usual definition of efolds number i.e.
$N=\ln{a_{end}\over a_{*}}=(1+{q\over 2})\mathcal{N}$. If $H$
changes insignificantly during inflation, like in the slow roll
and in the de Sitter inflation, $N\simeq \mathcal{N}$. By
substituting (\ref{11}) and (\ref{s9}) in (\ref{s20}) we arrive at
\begin{equation}\label{s21}
\mathcal{N}={q\over 2}\left({2\over 3\gamma}-1\right)\ln{\Phi_{*}\over \Phi_{end}},
\end{equation}
which in the high friction regime gives (see (\ref{s8}))
\begin{equation}\label{s21a}
\mathcal{N}=\ln{\Phi_{*}\over \Phi_{end}},
\end{equation}
while for the minimal case
\begin{equation}\label{s21b}
\mathcal{N}_{min}={1-q\over 3}\ln{\Phi_{*}\over \Phi_{end}},
\end{equation}
in agreement with \cite{Liddle}. By comparing these results, we
deduce that with a same ${\Phi_{*}\over \Phi_{end}}$ our model can
provide more e-folds than the minimal case. Note that in an
intermediate regime, where high friction condition does not hold,
obtaining an analytical solution for $a$ and $H$ is not feasible,
and we are unable to obtain a simple form for $\mathcal{N}$.

Now let us specify a lower bound for efold number during
inflation. Take $t_k$ as the time where a length   scale
$\lambda_k={1\over k}$, attributed to the wavenumber $k$, exited
the Hubble radius during inflation:
\begin{equation}\label{per1}
k={1\over \lambda_k}=a(t_k)H(t_k),
\end{equation}
where we have taken $a(t_0)=1$ and $t_0$ denotes the present time.
Large scale structure observations are limited to scales of about
1 Mpc (which we denote $\lambda_{minimum}$) to the present Hubble
radius (denoted by $\lambda_{max}$). These observable scales
crossed the Hubble radius during the following visible e-folding
(see (\ref{s20}))
\begin{equation}\label{Hubble}
\mathcal{N}_{vis}=\ln\left({\lambda_{max}\over
\lambda_{minimum}}\right)=\ln\left(H_0^{-1}\over 1Mpc\right).
\end{equation}
By inserting $H_0=67.3 km/s Mpc^{-1}$ \cite{Planck1} in
(\ref{Hubble}), we obtain $\mathcal{N}_{vis}=8.4$. Hence all
relevant scales exited the Hubble radius during $8.4$ e-folding
after  ${1\over H_0}$'s exit. Hence $\mathcal{N}>8.4$.

In the minimal case, from (\ref{s21b}) and
(\ref{s13}) we find
\begin{equation}\label{min1}
\mathcal{N}_{min}< {1-q\over 3}\ln\left(\left({\sqrt{3\over 8\pi}}
{q\Gamma\left( {q+2\over 2q}\right)\over
{\Gamma\left( {1\over q}\right)}}\right){M_P\over \Phi_{end}}\right)
\end{equation}
$\Phi_{end}$ depends on the chosen potential, e.g. for (\ref{s15})
with $d=1$, $\Phi_{end}\sim \phi_c$ \cite{Mukh}.  If we
take$\Phi_{end}$ of the same order as the electroweak scale,
$\Phi_{end}\sim 10^{-17}m_P$ (where $m_P$ is the planck mass), for
$q>0$, we obtain $\mathcal{N}_{min}<11.3$. By increasing the scale
of $\Phi_{end}$ this value decreases, for example for
$\Phi_{end}\sim 10^{-6}m_P$, and $q>0$,  we obtain
$\mathcal{N}_{min}<3.01$. Therefore in the minimal model
$\mathcal{N}_{vis}=8.4$ may be consistent with rapid oscillating
scalar phase and with the potential (\ref{s15}) provided that we
assume the extreme case (i.e. $\Phi_{end}$ is reduced to the
electroweak scale and $\Phi$ is augmented to the right hand side
of (\ref{s13})). In the nonminimal case, as the model is capable
to provide more e-folds than the minimal situation, the theory may
become more viable at least in the context of perturbations
generation.

Following (\ref{per1}), a wavenumber  had the possibility to exit
the Hubble radius during rapid oscillation phase, provided that the condition
$k\ll {1\over T(t_k)}$ (note that $a(t_0)=1$) holds. To study this condition, we proceed as follows: The largest scale of our observable
universe is of the same order of magnitude as
$\lambda_{max}={1\over H_0}$. Using (\ref{s11}) and (\ref{s13})
one can find an upper bound for $T$ during rapid oscillation:
$T(t)<T_u$. Hence $H_0 T_u\ll 1$ guarantees the compatibility of
our assumptions with the horizon exit of $\lambda_{max}$ during
rapid oscillation phase. This can be expressed as
\begin{equation}\label{exit}
H_0\ll{1\over  \sqrt{3}} \left({\sqrt{8\pi}\Gamma\left({1\over
q}\right)\over q\Gamma \left({q+2\over
2q}\right)}\right)^{-{q\over q+2}} \left(M_P^2M\lambda^{1\over
q}\right)^{q\over q+2}.
\end{equation}
If (\ref{exit}) holds and the model provided enough efolds
($\mathcal{N}>\mathcal{N}_{vis}$) after this exit, then we can
claim that other large cosmological observable scales had also the
possibility to exit the Hubble radius during this stage of
inflation. In the next part we will study perturbations generation
and, based on astrophysical data, find an estimation for
parameters of our model as well as for $\mathcal{N}$.

\section{Cosmological perturbations}
To study the scalar and the tensor fluctuations we decouple the
spacetime into two components,  the background and the
perturbation. The background is described by the homogeneous and
isotropic FLRW metric corresponding to the oscillatory inflation
in the context of non minimal derivative coupling model studied in
the previous section. To study quantum perturbations in rapid
oscillation stage we have to use the Mukhanov-Sasaki equation.
Mukhanov-Sasaki equation for scalar and tensor perturbations in
non-minimal derivative coupling model was obtained in
\cite{Germani2}
\begin{equation}\label{15}
{d^2v_{{(s,t)}k}\over d\eta^2}+\big(c_{s,t}^2k^2-{1\over
z_{s,t}}{d^2z_{s,t}\over d\eta^2}\big)v_{(s,t)k}=0.
\end{equation}
$c_s$ and $c_t$ are the sound speed for the scalar and the tensor
mode respectively and $k$ is wave number for mode function $v_k$.
The conformal time $\eta$ is defined by
\begin{equation}\label{12}
\eta(t)={\int}^t{dt'\over a(t')}
\end{equation}
and, $z_s$ and $ z_t$ are given by
\begin{equation}\label{16}
z_s=a(t){M_p\Gamma\over H}\sqrt{2\Sigma}\qquad
z_t=a(t)M_p{\sqrt{e_{ij}^\lambda e_{ij}^\lambda
}\over2}\sqrt{1-\alpha}.
\end{equation}
The polarization tensor is normalized to $e_{ij}^\lambda
e_{ij}^{\lambda^\prime}=2\delta_{\lambda\lambda^\prime} $.
$\Gamma$ and $\Sigma$ are defined as
\begin{equation}\label{17}
\Gamma={1-\alpha\over 1-3\alpha} \qquad
\Sigma=M^2\alpha\big[1+{3H^2\over M^2}\big({1+3\alpha\over
1-\alpha}\big)\big].
\end{equation}
In the above, $\alpha={\dot{\phi}^2\over 2M^2M_p^2}$ and $c_{s,t}$
is given by relation
\begin{equation}\label{19}
c_s^2={H^2\over\Gamma^2\Sigma}\varepsilon_s\qquad
c_t^2={1+\alpha\over 1-\alpha},
\end{equation}
where $\varepsilon_s$ is
\begin{equation}\label{20}
\varepsilon_s={1\over a(t)}{d\over dt}[{a(t)\Gamma\over
H}(1-\alpha)]-(1+\alpha).
\end{equation}

The equation (\ref{15}) was studied in the slow roll approximation
($\alpha\simeq 0$)  for quasi de-Sitter background in
\cite{Germani2}. From (\ref{14}) we find that $\alpha$ is nearly constant
\begin{equation}
\alpha\approx{\gamma\over2}={q\over3q+6}.
\end{equation}

By using (\ref{20}) and the  Raychaudhuri equation,
\begin{equation}\label{21}
-{\dot{H}\over H^2}(1-\alpha)={M^2\over
H^2}\alpha+3\alpha-{\dot{\alpha}\over H},
\end{equation}
we find that $\varepsilon_s$ in high friction limit becomes
\begin{equation}\label{22}
\varepsilon_s=-6\alpha\big({1-\alpha\over
1-3\alpha}\big)^2(1-{\dot{H}\over
H^2})+\alpha({-15\alpha^2-2\alpha+9\over (1-3\alpha)^2}).
\end{equation}
In the rapid oscillation stage $a(t)$ is a power law function of
time (see ({\ref{11})), therefore $\epsilon=-{\dot{H}\over
H^2}\approx{q\over q+2}$, which shows that $\varepsilon_s$ is
approximately a constant.

Using (\ref{17}) and (\ref{19}), one can show that $c_s$ becomes
\begin{equation}\label{23}
c_s^2\approx{(1-3\alpha)^2\over
3\alpha(1-\alpha)(1+3\alpha)}\varepsilon_s.
\end{equation}
Therefore,  $c_s$, is approximately a constant too . We can
calculate $c_s$ and $c_t$ as functions of $q$
\begin{equation}\label{24}
c_s^2\approx{q^3+8q^2+19q+18\over3(q+1)(q+2)(q+3)} \qquad
c_t^2={2q+3\over q+3}.
\end{equation}
In fig.(\ref{fig1}) $c_s$ is plotted with respect to $q$.
\begin{figure}[h]
\centering\epsfig{file=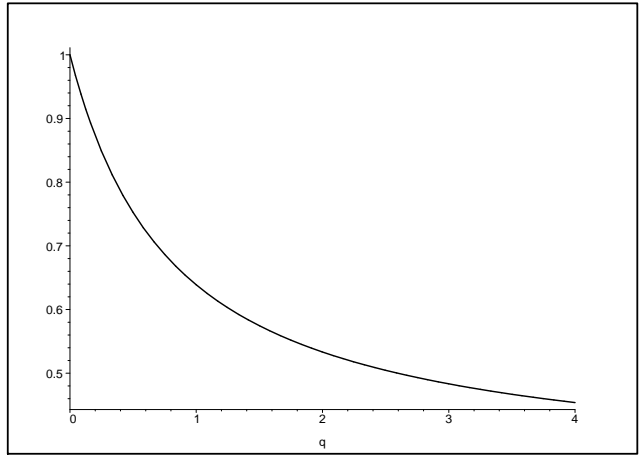,width=6cm,angle=270} \caption{$c_s$
in terms of $q$. } \label{fig1}
\end{figure}
This figure shows that sound speed for power law potentials is
restricted to the range $0<c_s<1$. $z$ in the rapid oscillation
era is
\begin{equation}\label{25}
z_s=a(t)M_p({1-\alpha\over1-3\alpha})\sqrt{6\alpha({1+3\alpha\over1-3\alpha})}.
\end{equation}
We have $a(\eta)\propto \eta^{-({q+2\over2})}$, thus we can write
$z$ in the form
\begin{equation}\label{26}
z_{s,t}=\beta_{s,t} a(\eta),
\end{equation}
where
\begin{equation}\label{27}
\beta_s\approx M_p{({q+3\over 3})}\sqrt{2q(q+1)\over q+2} \qquad
\beta_t=M_p{\sqrt{e_{ij}^\lambda e_{ij}^\lambda
}\over2}\sqrt{2q+6\over 3q+6}.
\end{equation}
So the conformal time derivative of $z$ is given by
\begin{equation}\label{28}
{1\over z_{s,t}}{d^2z_{s,t}\over
d\eta^2}=({q\over2}+1)({q\over2}+2)\eta^{-2}.
\end{equation}
Hence the mode function satisfies
\begin{equation}\label{29}
{d^2v_{(s,t)k}\over
d\eta^2}+\big(c_{s,t}^2k^2-({q\over2}+1)({q\over2}+2)\eta^{-2}\big)v_{(s,t)k}=0,
\end{equation}
whose solution is
\begin{equation}\label{30}
v_{(s,t)k}(\eta)=|\eta|^{1\over
2}[C_{s,t}^{(1)}(k)H_\nu^{(1)}(c_{s,t}k|\eta|)+C_{s,t}^{(2)}(k)H_\nu^{(2)}(c_{s,t}k|\eta|)].
\end{equation}
$C^{(1)}(k)$ and $C^{(2)}(k)$ are the constants of integration and
$H_\nu^{(1)}$ and $H_\nu^{(2)}$ are Hankle functions of the first
and second kind of order $\nu={3\over2}+{q\over2}$ respectively.
We adopt the Bunch-Davies vacuum by imposing the condition that
the mode function approaches the vacuum of the Minkowski spacetime
in the short wavelength limit ${a\over k}\ll{1\over H}$, where the
mode is well within the horizon. In the rapid oscillation epoch we
have $aH\propto{1\over |\eta|}$  resulting $k|\eta|\gg1$. In this
limit the Bunch-Davies mode function is given by $ v_k(\eta)
\approx {1\over \sqrt{2c_sk}}e^{-ic_sk\eta}$. This must be the
asymptotic form of (\ref{30}), therefore
\begin{equation}\label{31}
v_{(s,t)k}(\eta)={\sqrt{\pi}\over 2}e^{i(\nu+{1\over 2}){\pi\over
2}}{ (-\eta)}^{1\over 2}H_\nu^{(1)}(-c_{s,t}k\eta).
\end{equation}

In the limit ${k\over aH}\rightarrow0$ the asymptotic form of mode
function (\ref{31}) is given by
\begin{equation}\label{33}
v_{(s,t)k}(\eta)\rightarrow e^{i(\nu+{1\over 2}){\pi \over2 }
}2^{(\nu-{3\over2})}{\Gamma(\nu)\over\Gamma({3\over2})}{1\over\sqrt{2c_{s,t}k}}{(-c_{s,t}k\eta)}^{(-\nu+{1\over2})}.
\end{equation}

To obtain the power spectrum for scalar (tensor) perturbation we
follow the steps of \cite{kolb} and  substitute (\ref{33}) in
\begin{equation}\label{32}
{P_{s,t}(k)}^{1\over2}=\sqrt{{k^3\over 2\pi^2}}|{v_{(s,t)k}\over
z_{s,t}}|
\end{equation}
which yields
\begin{equation}\label{34}
P_{s,t}(k)^{1\over2}=\sqrt{k^3\over
2\pi^2}{2^{(\nu-{3\over2})}\over\beta_{s,t}
a}{\Gamma(\nu)\over\Gamma({3\over2})}{1\over\sqrt{2c_{s,t}
k}}{(-c_{s,t}k\eta)}^{(-\nu+{1\over2})}.
\end{equation}
We rewrite the conformal time as
\begin{equation}\label{35}
\eta=\int{{dt\over a(t)}}=\int{{da\over a^2H}}=-{1\over
aH}+\int{{\epsilon da\over a^2H}},
\end{equation}
but in the rapid oscillation epoch $\epsilon$ is a constant,
therefore
\begin{equation}\label{36}
\eta=-{1\over aH}{1\over 1-\epsilon}.
\end{equation}
By substituting (\ref{36}) into the equation (\ref{34}), we arrive
at
\begin{equation}\label{37}
P_{s,t}(k)^{1\over2}={2^{(\nu-{5\over2})}\over
\pi}{\Gamma(\nu)\over\Gamma({3\over2})}{k\over\sqrt{c_{s,t}}\beta_{s,t}
a}{({c_{s,t}k\over aH}{1\over 1-\epsilon})}^{(-\nu+{1\over2})}.
\end{equation}
At the horizon crossing $c_sk=aH$,
\begin{equation}\label{38}
P_{s,t}(k)^{1\over2}={2^{(\nu-{5\over2})}\over
\pi}{\Gamma(\nu)\over\Gamma({3\over2})}{H\over
c_{s,t}^{{3\over2}}\beta_{s,t}
}{(1-\epsilon)}^{(\nu-{1\over2})}.
\end{equation}
The above relation may be written as
\begin{equation}\label{39}
P_{s,t}(k)^{1\over2}=A_{s,t}(q){H\over M_p}|_{c_{s,t}k=aH},
\end{equation}
where
\begin{equation}\label{40}
A_s(q)={3^{{7\over4}}{2^{(q-{1\over2})}}{\Gamma({3\over2}+{q\over2})}(q+2)^{(-{q+1\over2})}\over
\pi\Gamma({3\over2})\sqrt{q(q+1)}(q+3)\left({q^3+8q^2+19q+18\over(q+1)(q+2)(q+3)}\right)^{{3\over4}}}
\end{equation}
and
\begin{equation}\label{40.5}
A_t(q)={3^{{1\over2}}{2^{(q-{1\over2})}}{\Gamma({3\over2}+{q\over2})}
(q+2)^{(-{q+1\over2})}\over\pi\Gamma({3\over2})(q+3)^{-{1\over4}}(2q+3)^{{3\over4}}}.
\end{equation}
The ratio of the tensor to scalar spectrum is given by
\begin{equation}\label{40.6}
r={P_t\over p_s}={({A_t\over A_s})}^2
={\sqrt{3}q(q+1)(q+3)^{{5\over2}}\over27(2q+3)^{{3\over2}}}\left({q^3+8q^2+19q+18\over(q+1)(q+2)(q+3)}\right)^{{3\over2}}.
\end{equation}

Now we can calculate the spectral index by differentiating the
power spectrum with respect to $k$ at horizon crossing
\begin{equation}\label{41}
n_s-1={d\ln{P_s} \over d\ln{k}}|_{c_sk=aH}.
\end{equation}
At the horizon crossing, we have ${d\ln{k}\over
dt}=H(1-\epsilon)$, so
\begin{equation}\label{42}
n_s-1={d\ln{H^2}\over d\ln{k}}=-{2\epsilon\over 1-\epsilon}=-q.
\end{equation}

Now, equipped with these results, we are capable to use
astrophysical data to fix the parameter $q$. For the pivot mode
$k_*=0.05 Mpc^{-1}$, the power spectrum and the spectral index are
determined from Planck 2013 data as  (for $\%68 CL$, or $1\sigma$
error)\cite{Planck1}
\begin{eqnarray}\label{47}
P_s&=&(2.200\pm0.056)\times10^{-9}\nonumber\\
n_s&=&0.9608\pm0.0054
\end{eqnarray}
Therefore (\ref{42}) leads to
\begin{equation}\label{s22}
q=0.0392\pm0.0054.
\end{equation}
For $q=0.0392$, the tensor scalar ratio is
\begin{equation}
r\approx0.0387,
\end{equation}
which is in agreement with Planck data which put an upper bound on $r$,
$r< 0.11 (95\% CL)$ \cite{Planck1}.

Now we are able to determine the range of the parameters required for
validity of the high friction and rapid oscillation assumptions.
$A_s(q=0.0392)=0.7993$ and
(\ref{39}) give the energy density of the scalar field at horizon
crossing as:
\begin{equation}\label{energy}
\rho_{*}\approx 1.032\times10^{-8}M_p^4\simeq 36.28
\times (10^{16} GeV)^4,
\end{equation}
which is compatible with the fact
that our model does not enter in the quantum gravity regime. Using
(\ref{s22}),  the rapid oscillation condition (\ref{s12}) reduces
to
\begin{equation}\label{s23}
\Phi_{*}^{1.019}\ll 0.0393{M_P^2 M\over \sqrt{\lambda}}.
\end{equation}
Where $\Phi_{*}$ is the scalar field amplitude at the horizon
crossing. In high friction regime one has ${H_*^2\over M^2}\gg 1$,
and in the rapid oscillation stage
$\lambda\Phi_{*}^{0.0392}=\rho_*=3M_P^2H_*^2$ holds. By collecting all these
results together we find
\begin{eqnarray}\label{s24}
&&\Phi_{*}\ll 386.8 M\nonumber \\
&&M^2\ll 0.344\times10^{-8}M_p^2,
\end{eqnarray}
and
\begin{equation}\label{s24a}
\tilde{\lambda}^{1\over 0.0392}\tilde{M}\gg {(1.032\times 10^{-8})^{1\over
0.0392}\over 386.8}
\end{equation}
where the dimensionless parameters $\tilde{M}$ and $\tilde{\lambda}$ are defined through $\lambda=\tilde{\lambda}M_P^{4-q}$,
and $M=\tilde{M} M_P$.
By inserting (\ref{s22}) and $H_0=67.3 km/s
Mpc^{-1}$ \cite{Planck1} in (\ref{exit}), we derive
\begin{equation}\label{exit1}
\tilde{\lambda}^{1\over
2.0392}\tilde{M}^{0.0392\over 2.0392}\gg 1.102\times 10^{-60}.
\end{equation}
Note that there is an interval of $5.4$
e-folds between the exits of $H_0$ and $k_*=0.05 Mpc^{-1}$, from
the Hubble radius: $ln\left({k_*\over H_0}\right)\simeq 5.4$. In our
computations we have assumed that the high friction condition is
still valid until the end of rapid oscillatory inflation, hence
$M^2\ll H_{end}^2$, which puts a stronger constraint on $M$
\begin{equation}\label{s24aa}
M^2\ll 0.344\times 10^{-8}e^{-0.0392\mathcal{N}}M_P^2.
\end{equation}
To derive (\ref{s24aa}),
\begin{equation}\label{s24aaa}
\mathcal{N}=\ln\left({\Phi_*\over \Phi_{end}}\right)={1\over q}\ln\left({\rho_*\over \rho_{end}}\right)
\end{equation}
was used. Note that $N\simeq 1.02 \mathcal{N}$, where
$N=\ln\left({a_{end}\over a_*}\right)$. The smallness of $q$ gives us the
option to choose $M$ (in (\ref{s24aa})) such that $\Lambda=(M^2
M_P)^{1\over 3}$ and consequently $\Lambda=(H^2 M_P)^{1\over 3}$ (
the cut-off scale during inflation
\cite{Germani1},\cite{Germani3}) become much larger than the TeV
scale.

The number of e-folds from the horizon crossing (of the pivot
mode $k_*$), until the end of inflation in the rapid oscillation
stage can be determined from (\ref{s21a}). Due to the smallness of
$q$, we may have ${\Phi_*\over \Phi_{end}}\gg 1 $, while $H_*$ and
$H_{end}$ have the same order of magnitude. For example for
$\mathcal{N}=60$, we have ${\Phi_*\over \Phi_{end}}=e^{60}$, while
$H_*=3.2 H_{end}$. This look likes like the slow roll situation
where $H$ decreases very slowly during inflation. The ratio
${\Phi_*\over \Phi_{end}}$ can not be fixed via our derived
relations. A formal upper bound for $\mathcal{N}$ may be extracted
from (\ref{s24}), $\mathcal{N}< \ln\left({386.8 M\over \Phi_{end}}\right)$.
The rapid oscillation condition puts an upper bound on the scalar
field amplitude (see (\ref{s12})). Therefore this condition cannot
be violated during the expansion (in contrast to the slow roll
conditions), hence the end of inflation and $\Phi_{end}$ may not
be determined in terms of the actual parameters of our scalar
field model with a power law potential as may be usually done in
the slow roll models. Besides,  to minimize the uncertainties in
the evaluation of $\mathcal{N}$, one needs to study the evolution
of the universe after the inflation specially the reheating era.
If we consider a prompt reheating, then the energy scale at the end
of inflation may be approximated as the reheating temperature
which must be less the GUT scale. In this situation by taking $\rho_{end}=(10^{16}GeV)^4$,
we obtain a lower bound for e-folds number as
$\mathcal{N}>91$.  $\mathcal{N}$ reduces by adopting larger
values for $\rho_{end}$. For example by setting $\{\lambda=1.76\times 10^{-8}M_P^{4-0.0392}, \,\, M=10^{-6}M_P, \,\ \Phi_{end}=10^{-32}M_P\}$,
which lie on the allowed domain for the rapid oscillation in high friction regime,
one obtains $\rho_{end}= 3.45\times (10^{16}GeV)^4=0.0981\times 10^{-8} M_P^4$ and $\mathcal{N}=60$.
In fig.(\ref{fig3}), $\mathcal{N}$ is depicted in terms of
$x:={\rho_{end}\over 10^{-8}M_P^4}$.
\begin{figure}[h]
\centering\epsfig{file=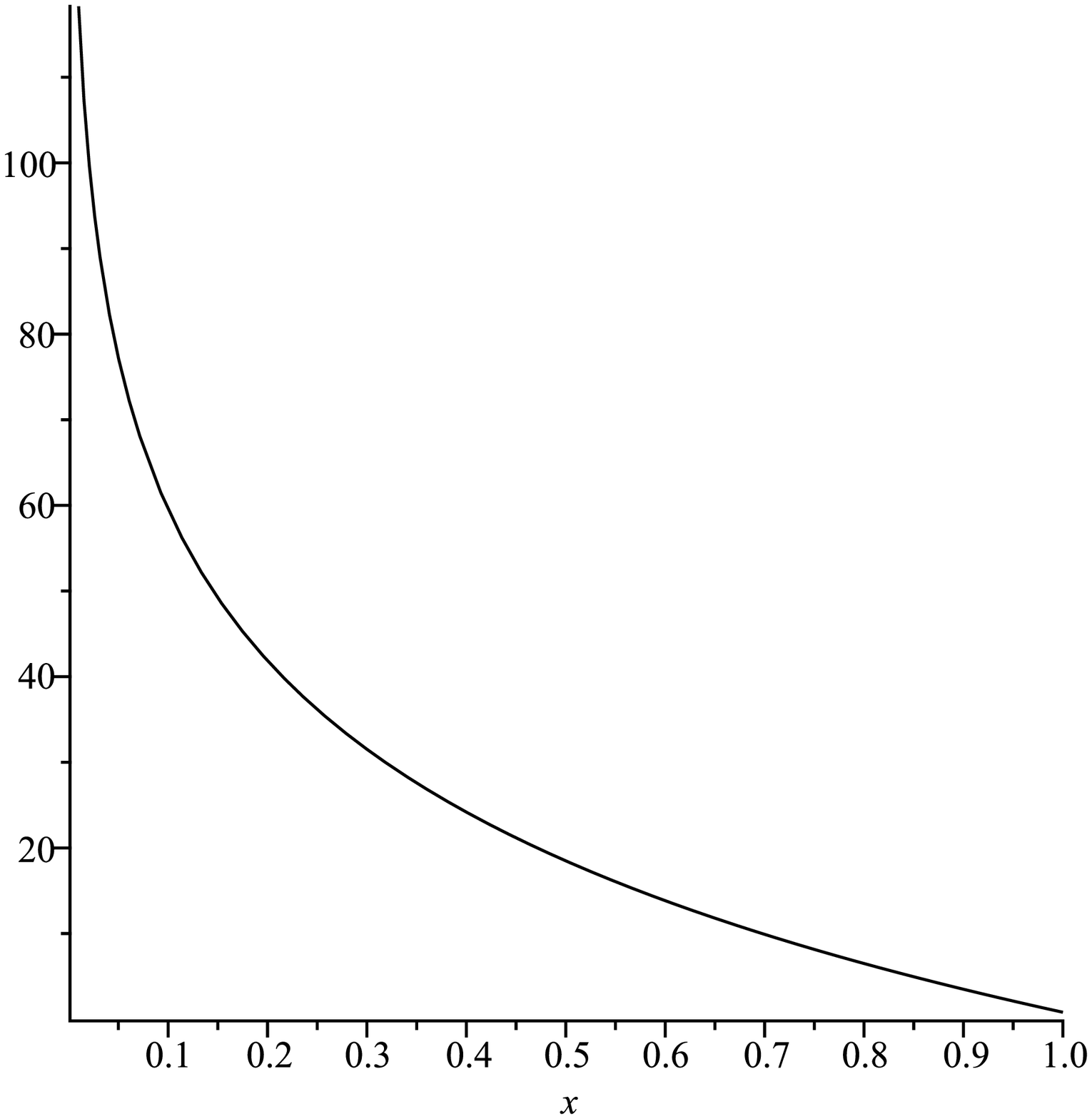,width=6cm,angle=0}
\caption{$\mathcal{N}$ in terms of $x:={\rho_{end}\over 10^{-8}M_P^4}$. }
\label{fig3}
\end{figure}
At the end let us note that if like \cite{Liddle}, one takes $\Phi_{end}\sim 5\times 10^{-17}M_P$,
and consider an extreme value for $\Phi_*$ , i.e. $\Phi_*=0.023M_P$ (derived from (\ref{s24})), the
number of e-folds from $t_{*}$ (time of horizon crossing) to
$t_{end}$  becomes $ \mathcal{N}=\ln{0.023 M_p\over {5\times 10^{-17}M_P}}= 35.4$,
which is approximately three times the number obtained in \cite{Liddle} and \cite{Mukh},
as is expected.

\section{conclusion}

Inflation driven by an oscillating scalar field with a power law
potential, $V(\phi)\propto \phi^q$, in the context of non minimal
derivative coupling model was studied. In high friction regime,
conditions required for such evolution were discussed. It was shown that,
in contrast to the
minimal case, $q$ is not restricted to a tighten limited range.
The number of e-folds, from a specific time (horizon crossing of a pivot scale)
in inflationary era until the end of inflation, was discussed.
Our results  indicate that more e-folds can be produced with respect to
the minimal case, giving the opportunity to observable cosmological scales
to exit the Hubble radius during inflation. Also, the conditions required for the
end of inflation were discussed.

We considered cosmological perturbations originated in this era and
computed  the power spectrum, the scalar spectral index, and the
tensor to scalar ratio.  By confronting our results with the
Planck 2013 data, we specified the range of the model parameters and
investigated the consistency of our results.


\begin{thebibliography}{99}
\bibitem{guth}A. H. Guth, Phys. Rev. D 23, 347 (1981);
A. Linde, Particle Physics and Inflationary Cosmology (Harwood,
Chur, Switzerland, 1990).
\bibitem{berz}F. L. Bezrukov and M. E. Shaposhnikov, Phys. Lett. B 659, 703 (2008);  F. Bezrukov,
A. Magnin, M. Shaposhnikov, and S. Sibiryakov, JHEP 01, 016
(2011).
\bibitem{Germani1}C. Germani and A. Kehagias, Phys. Rev. Lett. 105,
011302.
\bibitem{Sadj1}H. M. Sadjadi and P. Goodarzi , JCAP 02, 038,
(2013).
\bibitem{Sadj2}H. M. Sadjadi and P. Goodarzi , JCAP 07, 039,
(2013), arXiv:1302.1177[gr-qc].
\bibitem{Sadj3} H. M. Sadjadi, Phys. Rev. D 83,107301 (2011); E. N. Saridakis and S. V.
Sushkov, Phys. Rev. D 81, 083510 (2010); G. Gubitosi, and E. V.
Linder, Phys. Lett. B 703, 113 (2011); A. Banijamali and B.
Fazlpour, Phys. Lett. B 703, 366 (2011); S. V. Sushkov, Phys. Rev.
D 80, 103505 (2009).
\bibitem{infl}M. A. Skugoreva, S. V. Sushkov, and A. V.
Toporensky, arXiv:1306.5090 [gr-qc]; A. Ghalee, Phys. Lett. B 724,
198,(2013); G. Koutsoumbas, K. Ntrekis and E. Papantonopoulos
JCAP, 08, 027, (2013); K. Feng, T. Qiu, and Y. S. Piao,
arXiv:1307.7864v1 [hep-th]; R. N. Lerner and J. McDonald, Phys.
Rev. D 83, 123522 (2011).
\bibitem{Mukh}T. Damour and V. F. Mukhanov, Phys. Rev. Lett. 80, 3440, (1998).
\bibitem{Liddle}A. Liddle and A. Mazumder, Phys. Rev. D 58, 083508 (1996).
\bibitem{Planck1}P. A. R. Ade et al., Planck 2013 results. XVI, arXiv:1303.5076 [astro-ph]
(2013); P. A. R. Ade et al., Planck 2013 results. XXII,
arXiv:1303.5082 [astro-ph) (2013).
\bibitem{rapid}Y. Shtanov, J. Traschen, and R. Brandenberger, Phys. Rev. D 51, 5438
(1995).
\bibitem{kolb1} E. Kolb and M.Turner, The Early Universe (Addison-Wesley Publishing
Company, Redwood City, California, 1990).
\bibitem{Gu}J. A. Gu, arXiv:0711.3606 [astro-ph].
\bibitem{ref} Z. G. Liu, J.  Zhang, and Y. S. Piao, Phys.  Rev. D 84, 063508 (2011),
arXiv:1105.5713 [astro-ph.CO]
\bibitem{sami}M. Sami,  Grav. Cosmol. 8, 309 (2003),
arXiv:gr-qc/0106074v1.
\bibitem{Germani2}C. Germani and Y. Watanabe, JCAP 07(2011)031; C. Germani, L. Martucci, and P.
Moyassari, Phys. Rev. D, 85, 103501 (2012); C. Germani,
arXiv:1112.1083v1 [astro-ph.CO].
\bibitem{kolb}J. E. Lidsey, A. R. Liddle, E. W. Kolb, E. J.
Copeland, T. Barreiro, and M. Abneym Rev. Mod. Phys. 69, 2 (1997).
\bibitem{Germani3} S. Folkerts, C. Germani, and J. Redondo, arXiv:1304.7270v3
[hep-ph].





























\end{thebibliography}
\end{document}